# Optomagnetic forces on YIG/$YFeO_3$ microspheres levitated in chiral hollow-core photonic crystal fibre

SOUMYA CHAKRABORTY,[1,2] GORDON K. L. WONG,[2] FERDI ODA,[3,4] VANESSA WACHTER,[5,2] SILVIA VIOLA KUSMINSKIY,[5,2] TADAHIRO YOKOSAWA[6], SABINE HÜBNER,[6] BENJAMIN APELEO ZUBIRI,[6] ERDMANN SPIECKER,[6] MONICA DISTASO,[3,4] PHILIP ST. J. RUSSELL,[2] AND NICOLAS Y. JOLY[1,2]

*[1]Department of Physics, Friedrich-Alexander-Universität Erlangen-Nürnberg, 91058 Erlangen, Germany*
*[2]Max Planck Institute for the Science of Light, Staudtstraße 2, 91058 Erlangen, Germany*
*[3]Institute of Particle Technology, Friedrich-Alexander-Universität Erlangen-Nürnberg, Cauerstraße 4, 91058 Erlangen, Germany*
*[4]Interdisciplinary Center for Functional Particle Systems, Friedrich-Alexander-Universität Erlangen-Nürnberg, Haberstraße 9a, 91058 Erlangen, Germany*
*[5]Institute for Theoretical Solid-State Physics, RWTH Aachen University, 52074 Aachen, Germany*
*[6]Institute of Micro- and Nanostructure Research and Center for Nanoanalysis and Electron Microscopy, Friedrich-Alexander-Universität Erlangen-Nürnberg, Interdisciplinary Center for Nanostructured Films, Cauerstraße 3, 91058 Erlangen, Germany*
**Corresponding author: soumya.chakraborty@mpl.mpg.de / nicolas.joly@fau.de*

**We explore a magnetooptomechanical system consisting of a single magnetic microparticle optically levitated within the core of a helically twisted single-ring hollow-core photonic crystal fibre. We use newly-developed magnetic particles that have a core of antiferromagnetic yttrium-ortho-ferrite ($YFeO_3$) and a shell of ferrimagnetic YIG ($Y_3Fe_5O_{12}$) approximately 50 nm thick. Using a 632.8 nm probe beam, we observe optical-torque-induced rotation of the particle and rotation of the magnetization vector in presence of an external static magnetic field. This one-of-a-kind platform opens a path to novel investigations of optomagnetic physics with levitated magnetic particles.**

The optical tweezer technique has revolutionized our ability to trap and manipulate mesoscopic particles. [1–3] Magnetic fields provide an additional tool for manipulating tweezered magnetic particles. [4] Strong trapping of micro/nanoparticles in free-space requires a tightly-focused laser beam, whose depth of focus is limited by the Rayleigh length; to maintain a linear trap over long distances, the trapping beam must be both tightly focused and diffraction-free, which are conflicting requirements. [5] Hollow-core photonic crystal fibre (HC-PCF) uniquely provides a means of achieving a linear trap [5–8], light being confined to the hollow core either by a photonic bandgap [9] or by anti-resonant reflection. [10] The hollow core also provides a protected environment that can be adjusted through the addition of gases or liquids, or by evacuation. [11,12] This is an asset in studies of the effects of particle birefringence on the optical forces [13] and in more applied experiments such as living cell delivery [14] or thermal sensing using doped particles. [15]

In this paper we report optical trapping of magnetic microparticles and investigate their response to an external magnetic field. We focus on particles formed from yttrium-iron-garnet (YIG), which is a dielectric with a strong magnetic response. [16] Although it is transparent in the infrared, its relatively high refractive index (2.2 at 1064 nm) makes it challenging to trap optically. [17,18] The typical core diameter of HC-PCF is a few tens of µm, so that µm-sized particles can be conveniently accommodated. [5,7] Although it is possible to trap smaller (~100 nm-scale) particles, the overlap between light and particle is very small, making experiments difficult. Various techniques have been used to synthesize nm-scale YIG particles, such as co-precipitation, [19] sol-gel, [20] micro-emulsion, [21] microwave irradiation, [22] and traditional solid–state reaction methods. [23] Although current synthesis techniques permit a degree of control of particle size, they do not yet allow control of particle shape and surface roughness, which is irregular and unpredictable. This prevents the formation of high-Q internal optical resonances which are needed to enhance the weak photon-magnon coupling. [4]

The particles used in the experiments have a hybrid core-shell structure, the shell being a layer of cubic ferrimagnetic YIG ($Y_3Fe_5O_{12}$) approximately 50 nm thick, and the core a sphere of biaxial antiferromagnetic yttrium-ortho-ferrite ($YFeO_3$). They are propelled into a chiral [24] single-ring HC-PCF [25,26] using a dual-beam trapping scheme [7]. Since the hybrid particles are on average optically biaxial, they experience a torque when subject to a linearly polarized light beam, causing them to align along the electric field of the light. Subsequent application of a static external magnetic field results in anisotropic changes in magnetic permeability that in turn cause anisotropic changes in complex refractive index (the Voigt effect [27–29]) that are probed using HeNe laser light at 632.8 nm.

the proportion of the YIG phase increases (see Methods in SM). SEM characterization of particles isolated at 1000°C showed that the majority of them had a spherical shape with median diameter $x_{50,0}$ of 1.33±0.5 µm (Fig. 1b, c). Individual particles appear to have a rough and porous surface (Fig. 1b, c). We analysed particles calcined at 700°C by spin-coating

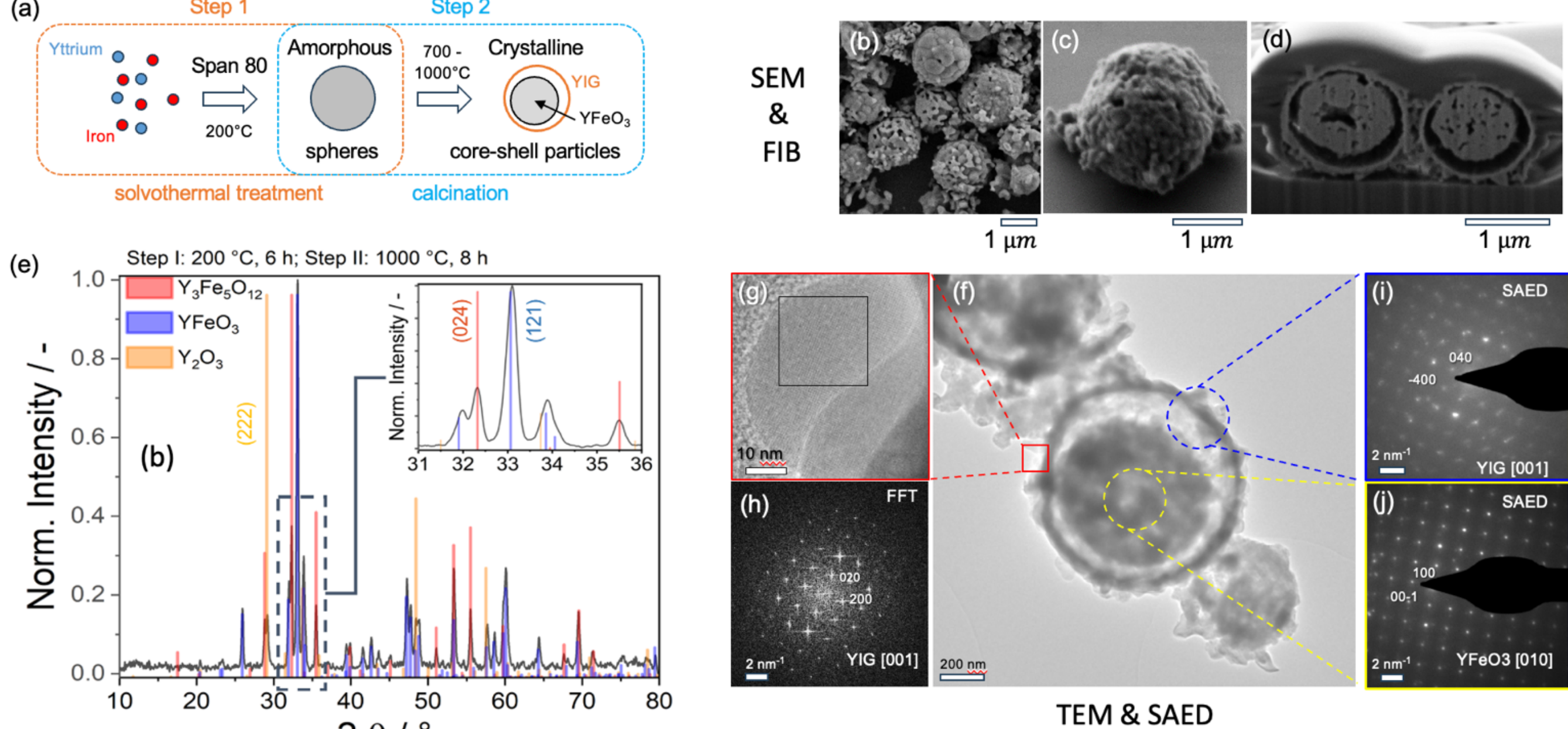

Fig. 1: Fabrication and characterization of magnetic YIG micron-sized particles. (a) Schematic of the synthesis of the particles. (b,c) SEM of the fabricated particles calcined at 1000°C and 700°C respectively, (d) FIB-cut of two particles calcined at 700°C revealing a consistent core-shell structure. (e) Powder x-ray diffraction (XRD) for particles fabricated at 1000°C revealing co-presence of three different crystalline materials. (f) HR-TEM image of a single particle calcined at 700°C. (g) High-resolution zoom of the shell region in the HR-TEM image and (h) the corresponding fast Fourier transform (FFT) revealing the cubic phase of YIG. (i) SAED pattern of the shell demonstrating cubic phase of YIG, (j) SAED of the core showing orthorhombic phase of $YFeO_3$.

Particle synthesis was carried out in two stages (Fig. 1a and methods in SM). First, the yttrium and iron molecular precursors were solubilized in N, N-dimethylformamide (DMF) and after the addition of the nonionic surfactant sorbitan monooleate- Span 80, the solution was aged under solvothermal conditions at 200°C for 6 hours. The surfactant Span 80 is well known to form micelles in polar solvents and has been used to synthesize porous alumina microparticles with interconnected pores, [30] $TiO_2$ microspheres, [31] gold nanoparticles, [32] $BaTiO_3$, [33] and to control the growth of $CaCO_3$. [34] The solid obtained was then isolated, washed, dried, resulting in amorphous spherical particles. In a second step the particles were calcined in air at temperatures between 700°C and 1000°C for 8 hours (Fig. 1a). The final particles were characterized by powder x-ray diffraction (XRD), scanning and transmission electron microscopy (SEM and TEM), high resolution-TEM, selected area electron diffraction (SAED) and energy-dispersive X-ray spectroscopy (EDS). In Fig. 1b powder diffraction of the particles calcined at 1000°C reveals the co-presence of three phases: orthorhombic $YFeO_3$ (ICSD 43260), cubic YIG (ICSD:14342), and minor quantities of cubic phase $Y_2O_3$ (ICSD:33648). It was observed that by increasing the calcination temperature, them on to a standard silicon wafer and sputtering them with 200 nm thick gold layer. We then cut them open using a Ga focused ion beam (FIB) in a Zeiss NVision 40. The FIB probe was set to 30 kV-80 pA. FIB-cut SEM images of the cut-open particles reveal a complex morphology (Fig. 1d). Their structure is egg-like, consisting of an external shell and a porous interior core. The composition of the different parts of the particles were analysed using EDX in conjunction with SAED and HR-TEM, revealing that the thin shell is made of YIG and the core contains $YFeO_3$ (Fig. 1g-n). There is no specific crystal orientation relationship between the shell and core.

In the optical experiments we trapped particles calcined at 700°C in a conventional dual-beam trap (Fig. 2). Light from a continuous-wave ytterbium fibre laser (Keopsys CYFL-KILO) delivering 3W at 1064 nm was used for the trapping beams. The light was split at a polarizing beam splitter (PBS) and coupled into the $LP_{01}$-like mode at both ends of a 7-cm-long chiral "single-ring" HC-PCF with core diameter 44 µm. This fibre has weak circular birefringence $B_C \sim 10^{-7}$ and is optically active, i.e., the electric field of a linearly polarized signal rotates slowly with distance while remaining linearly polarized, travelling around the equator of the Poincaré sphere. [24] Over the 7-cm length of the fibre this rotation is however very small so can be neglected. Precise preservation of linear polarization state is essential since we wish to probe small magnetically-induced anisotropic changes in complex dielectric constant. The fibre was mounted in a V-groove inside a custom-built low-pressure chamber with a transparent acrylic lid so as to allow access to light side-scattered by the particle as it is propelled along the fibre.

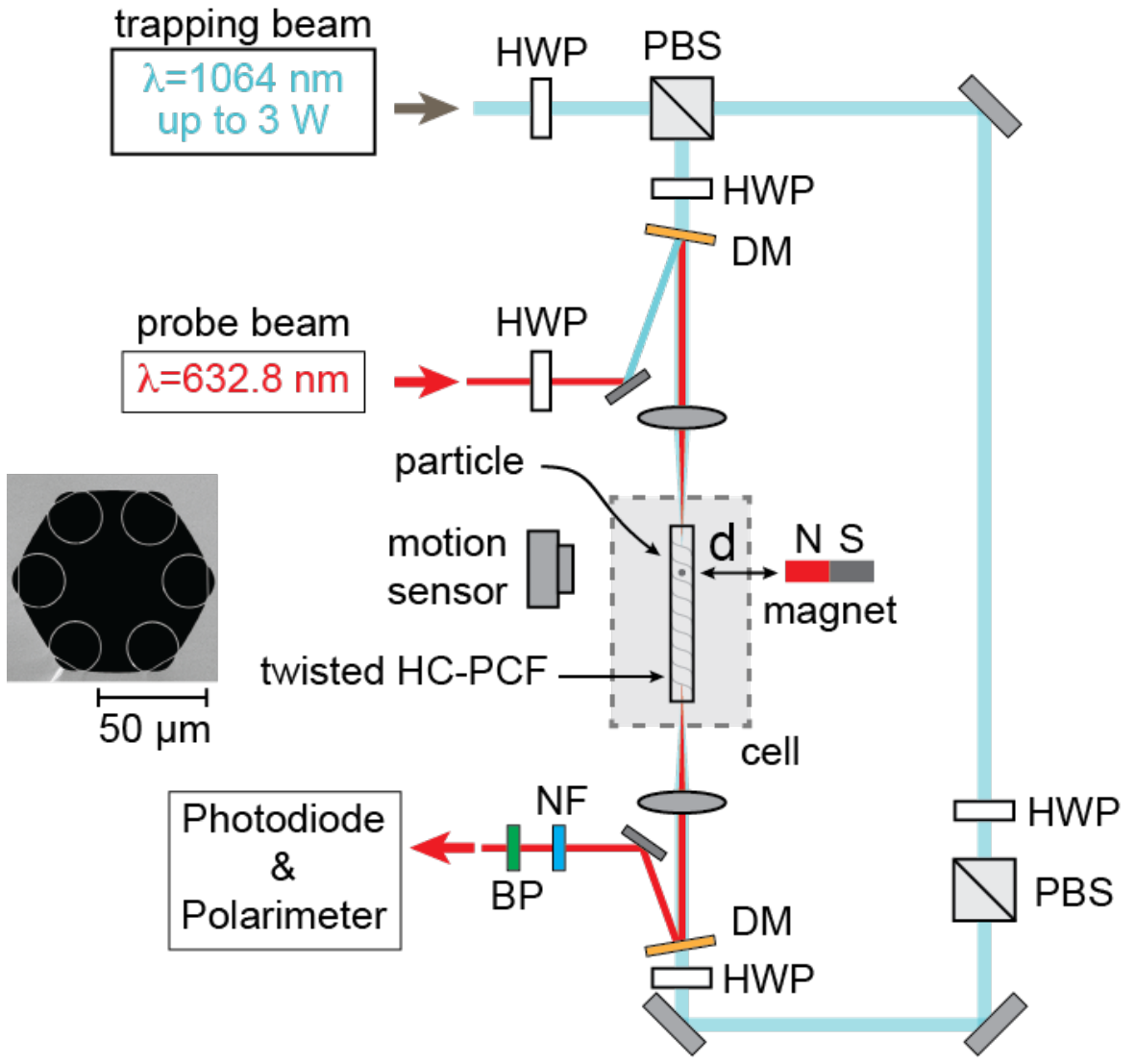

Fig. 2: Schematic of the experimental setup. HWP: half-wave plate, PBS: polarizing beam splitter, DM: dichroic mirror, NF: notch filter at 1064 nm, BP: 1 nm bandpass filter centred at 632.8 nm. The distance between the end of the magnet and the fibre is $d$. Inset: Scanning electron micrograph showing the cross-section of the single-ring HC-PCF. The core diameter of the fiber is 44 μm and the outer diameter ~270 μm. The motion sensor is a quadrant photodiode.

Particles were launched using the aerosol method. [7] The particles are first dispersed in a 50-50 mixture of a span-80 and water. A medical nebulizer was used to produce small aerosol droplets which were then delivered through an inlet placed above the fibre input face until one of the particles was trapped in front of the fibre. Once trapped at the entrance of the HC-PCF, a particle could be held there long-term and propelled into the core by momentarily lowering the power of the counter-propagating trapping beam. The trapping success rate with these particles was close to 100%. The motion of the bound particle along the fibre was imaged using a high-speed camera (Mikrotron EoSens mini2) placed above the chamber and a quadrant detector (Thorlabs PDQ30C) connected to an oscilloscope (PicoScope 3406B). A magnet was mounted on a translation stage so as to allow the magnetic field strength to be varied.

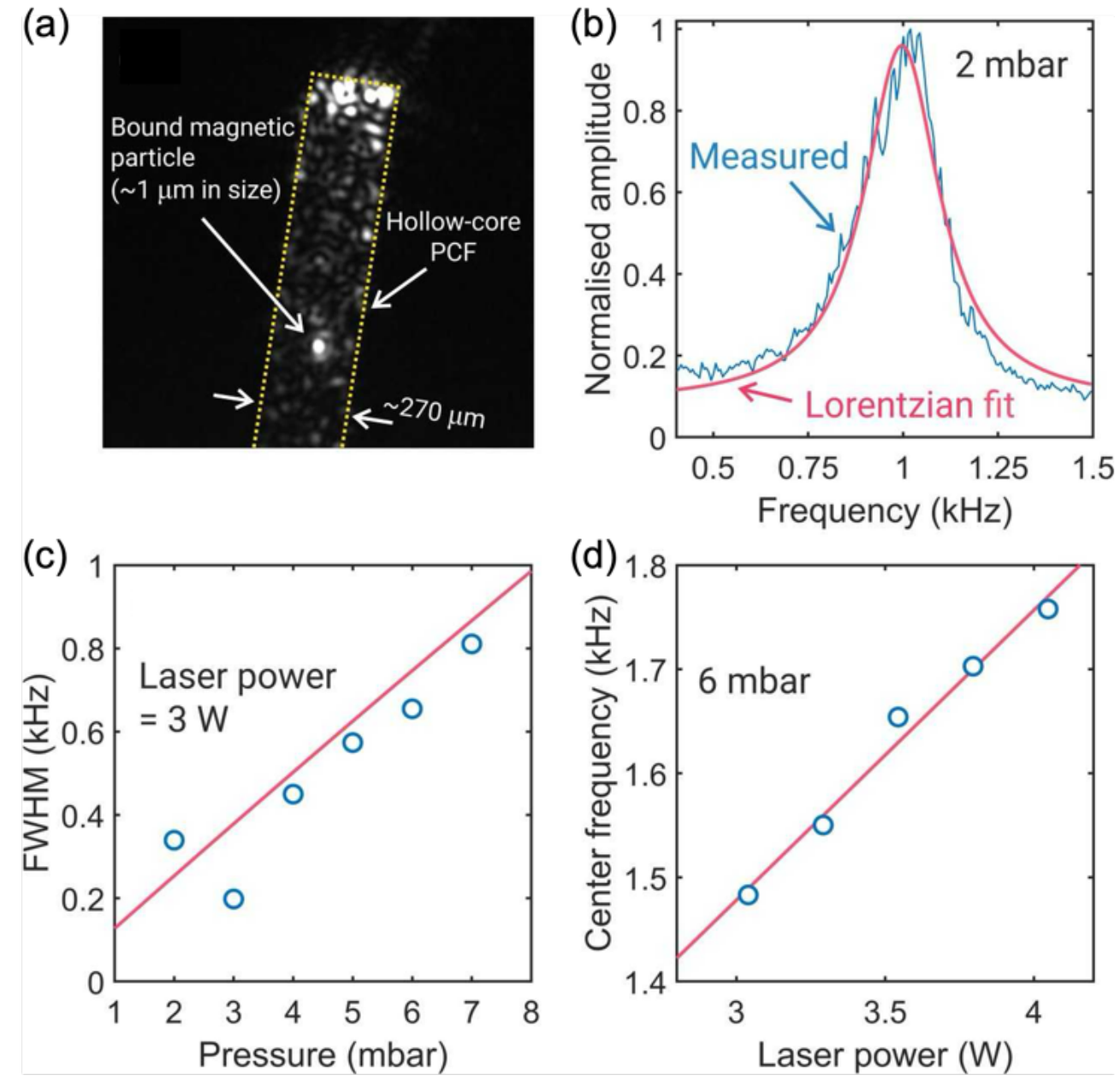

Fig. 3: (a) Snapshot of an optically trapped magnetic particle inside the core of a HC-PCF captured with a high-speed camera. (b) Spectrum of the damped mechanical motion of the bound magnetic particle at a pressure of 2 mbar. The laser power is 3 W. (c) Measured spectral linewidth (FWHM) as a function of the environment pressure at a fixed laser power of 3 W. The red line is the theoretical prediction according to Eq. (1). (d) Measured central frequency as a function of laser power at 6 mbar pressure.

First, we tested the limits of the levitated system by evacuating the chamber with a particle already trapped in the fibre core, as shown in Fig. 3a. At pressures below ~1 mbar the particle escaped from the optical trap and was lost, which we attribute to the onset of the ballistic regime caused by the increased molecular mean-free path. [35] In the transverse direction the trapped particle behaves like a damped mechanical oscillator, driven by Brownian motion. [12] From time-domain data recorded with the position-sensitive quadrant detector we extracted the Lorentzian spectrum of the particle motion (Fig. 3b). As the gas pressure decreases, the viscosity falls, and the linewidth narrows. The relationship between the spectral linewidth Γ and the air pressure $p$ follows the Knudsen relation:

$$\Gamma = \frac{\gamma}{m} = \frac{12\pi R}{m}\eta_0 \Big/ \left(1 + K_n(p)\left(\beta_1 + \beta_2 e^{-\beta_3/K_n(p)}\right)\right) \qquad (1)$$

where $\gamma$ is the damping coefficient caused by viscosity, $2R$ is a characteristic length (the particle diameter), and $m$ is the particle mass (densities of YIG and $YFeO_3$ are 5.11 and 5.47 $g/cm^3$ respectively) and $K_n(p) = 66 \times 10^{-9}/(pR)$ is the Knudsen number, $\eta_0 = 18.1$ μPa·s is the viscosity of air at atmospheric pressure and $\beta_1 = 1.231$, $\beta_2 = 0.469$ and $\beta_3 = 1.178$ are dimensionless constants. [7]

Figure 3(c) plots the measured spectral linewidth as a function of pressure, and the red line is a linear fit to Eq. (1). The trap stiffness, which governs the resonant frequency, is controlled by the trapping laser power $P$. At a fixed pressure (6 mbar in the experiment) the resonant frequency increases sublinearly with the laser power ($\Omega \propto \sqrt{P}$), as expected (Fig. 3d).

Crystalline $YFeO_3$ is orthorhombic and biaxial, displaying optical birefringence, which means that the linearly polarized trapping beam can be used as an optical spanner, [13] permitting measurements to be made as a function of particle orientation. The shell of the particles is formed from YIG, which is cubic and isotropic, becoming optically biaxial when a magnetic field is applied parallel to the (110) crystallographic plane (for details refer to SM). Both crystals are strongly absorbing at 632.8 nm, offering a simple means of probing magnetically induced changes in complex refractive index by monitoring the power and polarization state of the transmitted probe light [27,29,36].

The on-axis magnetic flux of the NdFeB permanent magnet (N35, cross-section 4×4 mm) used in the experiments is plotted in Fig. 6 as a function of the distance from the magnet's end-face. The magnet was placed with its N-S axis oriented perpendicular to the fibre axis and centred on the trapped particle (Fig. 1), and a motorized translation stage was used to vary the distance $d$ between the magnet and the particle. Probe light was provided by a linearly polarized HeNe laser (2 mW, 632.8 nm). The transmitted trapping beam light was filtered out using a combination of dichroic mirror, 1 nm bandpass filter centred at 632.8 nm, and 2 nm notch filter centred at 1064 nm (Fig. 1). In the experiments, both the power and the polarization state of the probe beam was monitored.

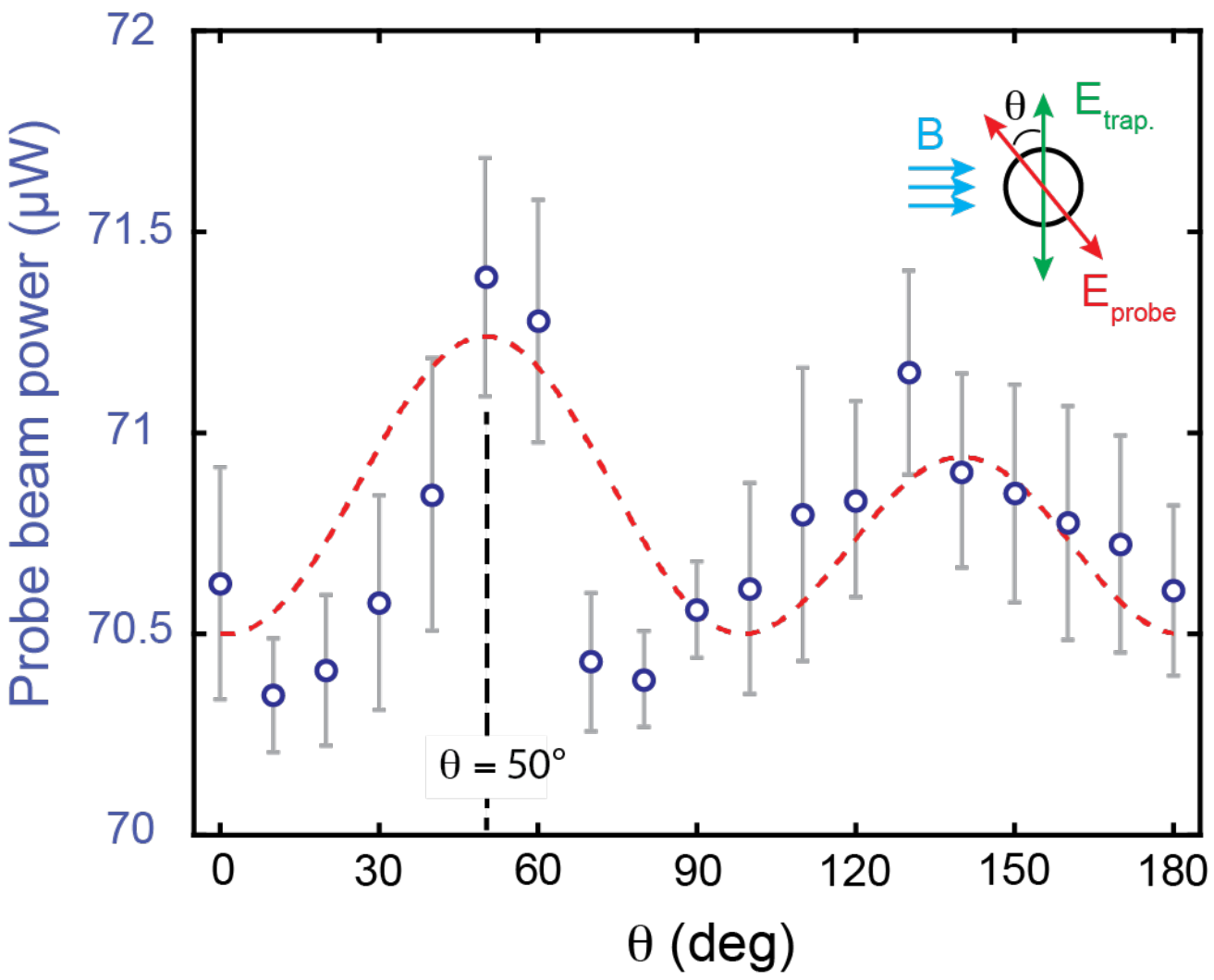

Fig. 4: Transmitted probe power as the particle is rotated inside the HC-PCF and subject to a constant magnetic field of 29 mT. The red dotted curve is a heuristic fit to Eq. (3).

The opto-magnetic response was first investigated by applying a constant magnetic field ($B$ = 29 mT) and rotating the particle by rotating the trapping beam polarization [13]. The transmitted probe power was directly monitored using both a polarimeter and a lock-in amplifier. Figure 4 plots the transmitted probe power as a function of the orientation of the trapping electric field $\theta$, where $\theta = 0°$ when the trapping and probe beams are co-polarized. For each value of $\theta$ we made repeated measurements of the transmitted power and evaluated the mean (blue dot) and standard deviation (error bar). Over 180° the data shows two distinct peaks, which we attribute to the complex-valued biaxial refractive index of the particle.

In the case of pure YIG crystal, assuming its magnetization vector points in the (110) plane, the imaginary part of the dielectric susceptibility can be written in the form [27,28]:

$$\Delta\chi_i(\theta) = \frac{M_S^2}{n_r}\left[g_{44} + \frac{\Delta g}{16}(3 + 2\cos 2\theta + 3\cos 4\theta)\right] \quad (2)$$

where $\theta$ is angle between the magnetization vector $\vec{M}$ and [001] crystal axis (see SM for detail), $n_r$ is the wavelength-dependent real part of the refractive index of YIG, $M_s$ is its saturation magnetization, $\Delta g = g_{11} - g_{12} - 2g_{44}$, and $g_{11}, g_{12}$ and $g_{44}$ are complex numbers representing the non-vanishing tensor elements of the dielectric tensor induced by a magnetic field and causing biaxial linear birefringence and (in the visible) dichroism.

Since in our case the particle is a complex hybrid of YIG and $YFeO_3$, the system cannot be so easily modelled. Moreover, when a new particle is launched into the trap, the initial orientations of its optical axes as well as the magnetisation axis are unknown. However, a heuristic fit to the power measurement can be made using a similar function with an added phase shift of $\psi$:

$$P(\theta) \propto \mathrm{a}\,[1 + \mathrm{b}\cos 2(\theta - \psi) + \mathrm{c}\cos 4(\theta - \psi)] \quad (3)$$

where the phase $\psi = -50°$ is added to the angle $\theta$, which in our experiment is the angle between probe and pump beam polarization. We note that $a$ represents the average power of our dataset which is 70.8 μW. The other coefficients are respectively $b = 2.12 \times 10^{-3}$ and $c = 4.1 \times 10^{-3}$. The Eq. (3) qualitatively fits to the data, as seen in the red dashed curve in Fig. 4. The individual magnetooptomechanic response of each particle is slightly different due to its initial orientation, though they all follow the same general trend, exhibiting two maxima (Fig. 4). The first peak occurs at $\theta \simeq 50°$ (Fig. 4), which is in reasonable agreement with the values for pure YIG (50°) and $YFeO_3$ (45°) (see SM for more details). [37]

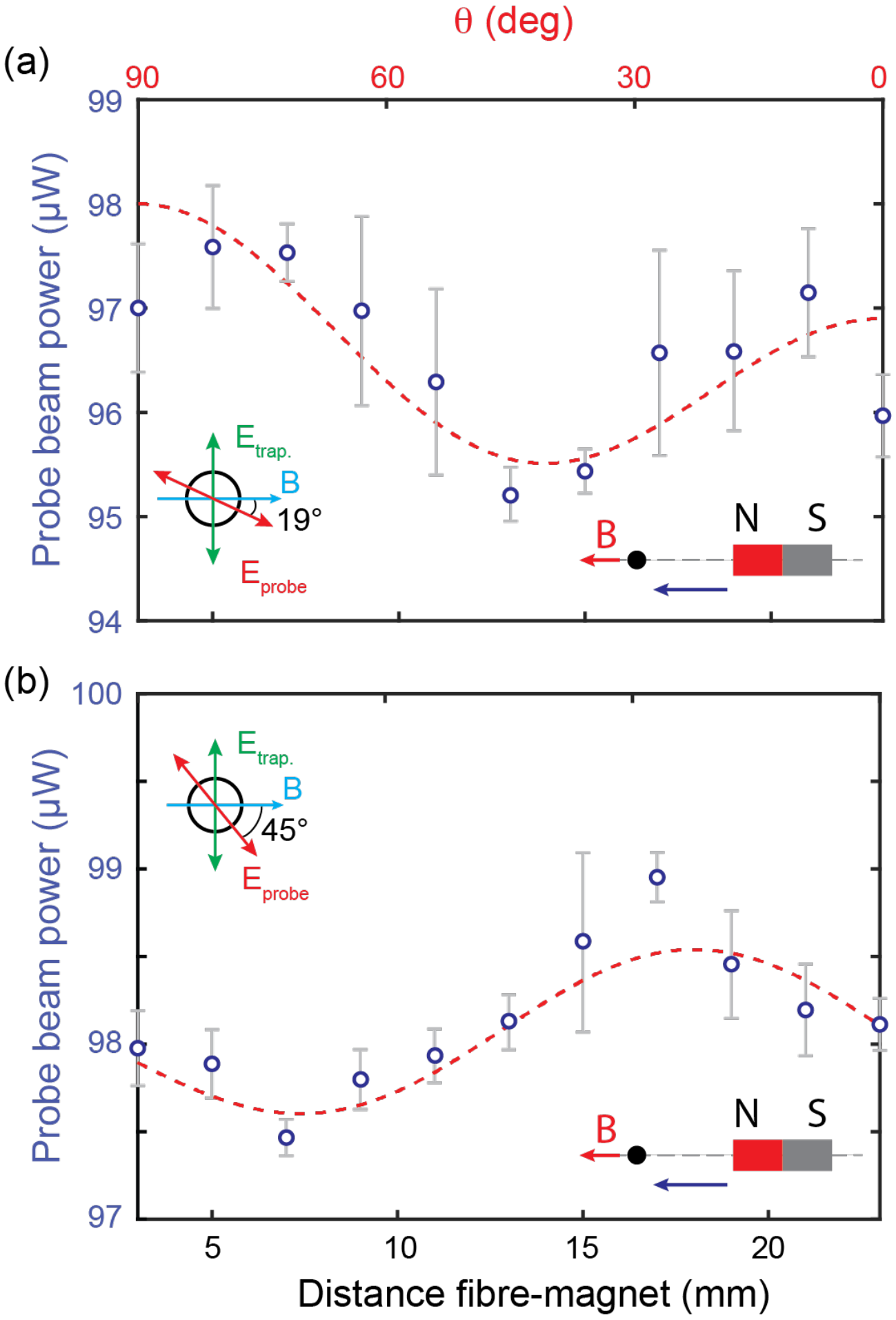

Fig. 5: Transmitted power (blue axis) as the magnet is moved inwards towards the particle. The angle between the orientation of the linearly polarized probe beam and the applied external field B is 19° in (a) and 45° in (b). The red dashed line is a heuristic fitting using Eq. (3)

Next, we kept the particle stationary and moved the magnet inwards, keeping the angle between the magnetic field and the probe beam polarization ($E_{probe}$) fixed to either 19° or 45°(Fig. 5 inset). In both cases, the magnetic field is orthogonal to the polarization of the trapping beam (Fig. 2). At $d = 23$ mm, the magnetic field $\vec{B}$ at the particle is very weak and the magnetization vector $\vec{M}$ is unaffected. As the magnet approaches the particle, $\vec{M}$ gradually rotates until it aligns almost parallel to $\vec{B}$ at $d = 3$ mm. At present, we cannot distinguish between the physical rotation of the particle from the rotation of only the magnetization vector, however they will give rise to same experimental result.
Figure 5(a) shows the variation of the transmitted probe beam power with respect to $d$ when the angle between magnetic field and the probe beam polarisation is 19° and Fig. 5(b) shows the variation when this angle is 45° as shown in the insets of the figures. The angle $\theta$ by which magnetization vector rotates depends inversely on $d$ allowing us to heuristically fit the experimental data with $P(\theta)$. For fitting of the data in Fig. 5(a), the added phase shift $\psi$ to $P(\theta)$ in Eq. (3) is $\psi = 90°$ and $a = 96$ μW, $b = 5.73 \times 10^{-3}$ and $c = 0.098$. For Fig. 5(b), the added phase shift is $\psi = -22.5°$ and the parameters $a = 98$ μW, $b = 1.53 \times 10^{-3}$ and $c = 3.98 \times 10^{-3}$ respectively. These plots show a behaviour similar to that in Fig. 4, from which we deduce that the imaginary part of the dielectric susceptibility is being probed as a function of the rotation angle $\theta$ and distance $d$. We also observed that once the external magnetic field was strong enough to align the magnetization vector parallel to itself, the transmitted probe power and polarisation state no longer responded to changes in magnetic field strength (see SM). A similar response could be recovered by moving the particle along the fibre out of the magnetic field and then returning it.

In summary, spheroidal μm-sized magnetic particles with a ~50 nm shell of YIG and a core of $FeO_3$ were synthesized. The relative proportion of the two materials could be adjusted by running the calcination process at different temperatures. The particles could be reproducibly trapped long-term in the evacuated HC-PCF core. Measurements with a 632.8 nm probe beam and a single μm-diameter particle reveal detectible changes in the transmitted power and the polarization state. The system is suitable for a wide variety of different applications, such as remote magnetic field sensing [36], interactions between waveguide modes, and the study of rotational degrees of freedom and spin waves in optomechanically cooled resonators [38]. The reported results suggest new possibilities for experiments in particle-based magneto-optomechanical physics., including cooling down to the single quantum regime [38], possibly at room temperature.

## Methods

The particles were fabricated in two steps. At first, an yttrium molecular precursor ($Y(NO_3)_3{\cdot}6H_2O$) (1.15 g, 3 mmol) and an iron molecular precursor ($Fe(acac)_3$) (1.8 g, 5.10 mmol) were solubilized at room temperature in 50 mL N,N-dimethylformamide (DMF) and the surfactant sorbitan monooleate - Span 80 was added during stirring. The as-obtained solution was transferred in a Teflon liner and aged in a stainless-steel autoclave at 200°C for 6 hours. Upon cooling to room temperature, toluene was added to the reaction mixture to induce precipitation. The solid was isolated by centrifugation and washed by three redispersion and centrifugation cycles. Finally, it was dried in air at 60°C for 20 hours. The product at this stage comprised amorphous spherical particles. The second step of the fabrication procedure comprised a calcination process in air at temperature between 700 and 1000 °C for 8 hours. This second part leads to an amorphous to crystalline phase transition during which the particles become magnetic.

The magnetic field was produced by eight $4 \times 4 \times 3$ mm$^3$ NdFeB N35 permanent magnets placed in a row. The on-axis magnetic flux density was measured with a Gaussmeter as a function of $d$, the distance from the end-face of the magnet (Fig. 6).

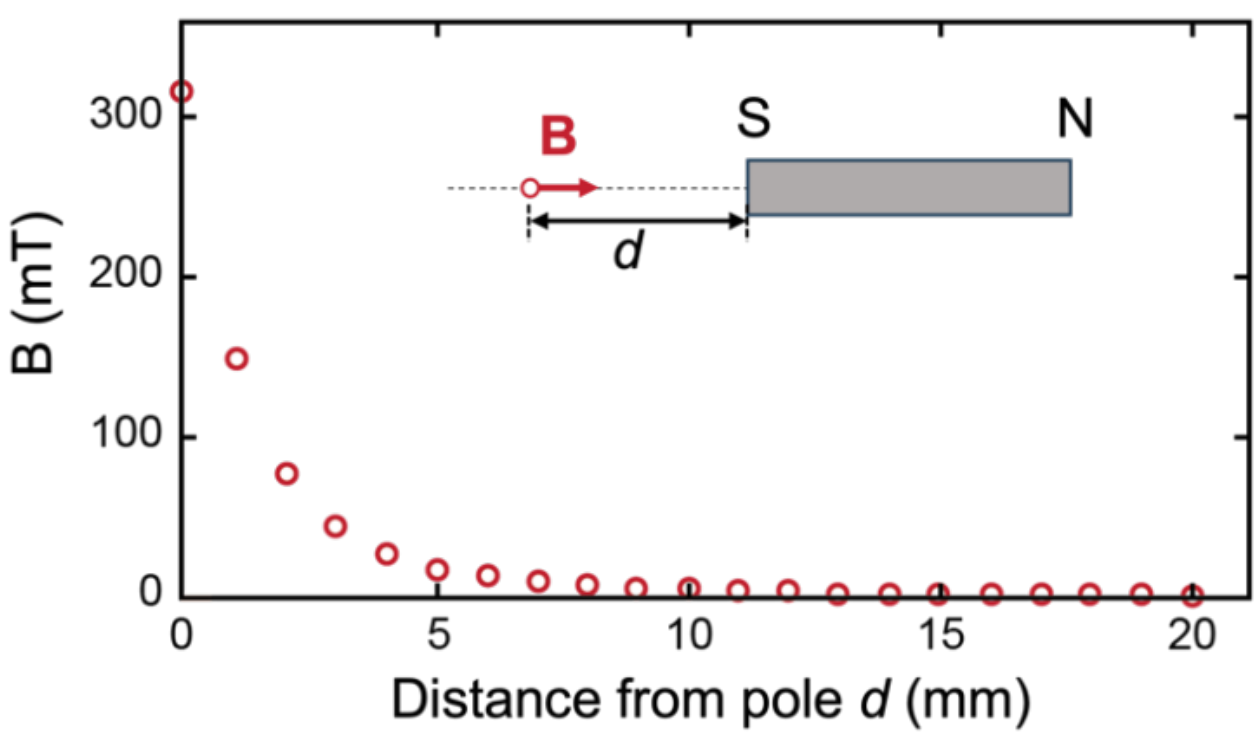

Fig. 6: Measured magnetic flux density **B** (mT) of the NdFeB magnet as a function of distance, along the N-S axis, from the one of the poles.

## References

1. A. Ashkin, "Acceleration and Trapping of Particles by Radiation Pressure," Phys. Rev. Lett. **24**, 156–159 (1970).
2. A. Ashkin, J. M. Dziedzic, and T. Yamane, "Optical trapping and manipulation of single cells using infrared laser beams," Nature **330**, 769–771 (1987).
3. I. A. Favre-Bulle, A. B. Stilgoe, E. K. Scott, and H. Rubinsztein-Dunlop, "Optical trapping in vivo: theory, practice, and applications," Nanophotonics **8**, 1023–1040 (2019).
4. V. Wachter, V. A. S. V. Bittencourt, S. Xie, S. Sharma, N. Joly, P. S. J. Russell, F. Marquardt, and S. V. Kusminskiy, "Optical signatures of the coupled spin-mechanics of a levitated magnetic microparticle," J. Opt. Soc. Am. B **38**, 3858–3871 (2021).
5. F. Benabid, J. C. Knight, and P. St. J. Russell, "Particle levitation and guidance in hollow-core photonic crystal fiber," Opt. Express **10**, 1195–1203 (2002).
6. M. J. Renn, D. Montgomery, O. Vdovin, D. Z. Anderson, C. E. Wieman, and E. A. Cornell, "Laser-Guided Atoms in Hollow-Core Optical Fibers," Phys. Rev. Lett. **75**, 3253–3256 (1995).
7. D. S. Bykov, O. A. Schmidt, T. G. Euser, and P. St. J. Russell, "Flying particle sensors in hollow-core photonic crystal fibre," Nat. Photonics **9**, 461–465 (2015).
8. D. S. Bykov, S. Xie, R. Zeltner, A. Machnev, G. K. L. Wong, T. G. Euser, and P. St. J. Russell, "Long-range optical trapping and binding of microparticles in hollow-core photonic crystal fibre," Light Sci Appl **7**, 22 (2018).
9. R. F. Cregan, B. J. Mangan, J. C. Knight, T. A. Birks, P. St. J. Russell, P. J. Roberts, and D. C. Allan, "Single-Mode Photonic Band Gap Guidance of Light in Air," Science **285**, 1537–1539 (1999).
10. C. Wei, R. J. Weiblen, C. R. Menyuk, and J. Hu, "Negative curvature fibers," Adv. Opt. Photon. **9**, 504–561 (2017).
11. R. Zeltner, D. S. Bykov, S. Xie, T. G. Euser, and P. St. J. Russell, "Fluorescence-based remote irradiation sensor in liquid-filled hollow-core photonic crystal fiber," Appl. Phys. Lett. **108**, 231107 (2016).
12. A. Sharma, S. Xie, and P. S. J. Russell, "Reconfigurable millimeter-range optical binding of dielectric microparticles in hollow-core photonic crystal fiber," Opt. Lett. **46**, 3909–3912 (2021).
13. S. Xie, A. Sharma, M. Romodina, N. Y. Joly, and P. St. J. Russell, "Tumbling and anomalous alignment of optically levitated anisotropic microparticles in chiral hollow-core photonic crystal fiber," Sci. Adv. **7**, eabf6053 (2021).
14. S. Unterkofler, M. K. Garbos, T. G. Euser, and P. St J Russell, "Long-distance laser propulsion and deformation-monitoring of cells in optofluidic photonic crystal fiber," J. Biophoton. **6**, 743–752 (2013).
15. J. Freitag, M. Koeppel, M. N. Romodina, N. Y. Joly, and B. Schmauss, "Flying Particle Thermosensor in Hollow-Core Fiber Based on Fluorescence Lifetime Measurements," IEEE J. Sel. Top. Quantum Electron. 1–9 (2023).
16. D. Vojna, O. Slezák, R. Yasuhara, H. Furuse, A. Lucianetti, and T. Mocek, "Faraday Rotation of $Dy_2O_3$, $CeF_3$ and $Y_3Fe_5O_{12}$ at the Mid-Infrared Wavelengths," Materials **13**, 5324 (2020).
17. "Deltronic Crystal Isowave," https://isowave.com/.
18. T. Seberson, P. Ju, J. Ahn, J. Bang, T. Li, and F. Robicheaux, "Simulation of sympathetic cooling an optically levitated magnetic nanoparticle via coupling to a cold atomic gas," J. Opt. Soc. Am. B **37**, 3714 (2020).
19. M. Rajendran, S. Deka, P. A. Joy, and A. K. Bhattacharya, "Size-dependent magnetic properties of nanocrystalline yttrium iron garnet powders," J. Magn. Magn. Mater. **301**, 212–219 (2006).
20. P. Vaqueiro, M. P. Crosnier-Lopez, and M. A. López-Quintela, "Synthesis and Characterization of Yttrium Iron Garnet Nanoparticles," J. Solid State Chem. **126**, 161–168 (1996).
21. P. Vaqueiro, M. A. López-Quintela, and J. Rivas, "Synthesis of yttrium iron garnet nanoparticlesviacoprecipitation in microemulsion," J. Mater. Chem. **7**, 501–504 (1997).
22. J. Liu, P. Yu, Q. Jin, C. Zhang, M. Zhang, and V. G. Harris, "Microwave-accelerated rapid synthesis of high-quality yttrium iron garnet nano powders with improved magnetic properties," Mat. Res. Lett. **6**, 36–40 (2018).
23. F. Chen, Z. Zhang, X. Wang, J. Ouyang, Z. Feng, Z. Su, Y. Chen, and V. G. Harris, "Room temperature magnetoelectric effect of $YFeO_3 – Y_3Fe_5O_{12}$ ferrite composites," J. Alloys Compd. **656**, 465–469 (2016).
24. P. St. J. Russell, R. Beravat, and G. K. L. Wong, "Helically twisted photonic crystal fibres," Philosophical Transactions of the Royal Society A: Mathematical, Physical and Engineering Sciences **375**, 20150440 (2017).
25. N. N. Edavalath, M. C. Günendi, R. Beravat, G. K. L. Wong, M. H. Frosz, J.-M. Ménard, and P. St.J. Russell, "Higher-order mode suppression in twisted single-ring hollow-core photonic crystal fibers," Opt. Lett. **42**, 2074 (2017).
26. J. C. Knight, J. Broeng, T. A. Birks, and P. St. J. Russell, "Photonic Band Gap Guidance in Optical Fibers," Science **282**, 1476–1478 (1998).
27. R. V. Pisarev, G. Sinii, N. N. Kolpakova, and Y. M. Yakovlev, "Magnetic birefringence of light in iron garnets," Sov. Phys.-JETP **33**, 1175–1182 (1971).
28. J. Ferre and G. A. Gehring, "Linear optical birefringence of magnetic crystals," Rep. Prog. Phys. **47**, 513 (1984).
29. W. Wettling, "Magnetooptical properties of YIG measured on a continuously working spectrometer," Appl. Phys. **6**, 367–372 (1975).
30. H. Yang, Y. Xie, G. Hao, W. Cai, and X. Guo, "Preparation of porous alumina microspheres via an oil-in-water emulsion method accompanied by a sol–gel process," New J. Chem. **40**, 589–595 (2016).

31. Z. Li, M. Kawashita, and M. Doi, "Sol–gel synthesis and characterization of magnetic $TiO_2$ microspheres," J. Ceram. Soc. Jpn **118**, 467–473 (2010).
32. C.-L. Chiang, "Controlled Growth of Gold Nanoparticles in Aerosol-OT/Sorbitan Monooleate/Isooctane Mixed Reverse Micelles," J. Colloid Interface Sci. **230**, 60–66 (2000).
33. R. Savo, A. Morandi, J. S. Müller, F. Kaufmann, F. Timpu, M. Reig Escalé, M. Zanini, L. Isa, and R. Grange, "Broadband Mie driven random quasi-phase-matching," Nat. Photonics **14**, 740–747 (2020).
34. C. Zhang, J. Zhang, X. Feng, W. Li, Y. Zhao, and B. Han, "Influence of surfactants on the morphologies of $CaCO_3$ by carbonation route with compressed $CO_2$," Colloids Surf. A: Physicochem. Eng. Asp. **324**, 167–170 (2008).
35. T. Li, S. Kheifets, D. Medellin, and M. G. Raizen, "Measurement of the Instantaneous Velocity of a Brownian Particle," Science **328**, 1673–1675 (2010).
36. O. Kamada, T. Nakaya, and S. Higuchi, "Magnetic field optical sensors using Ce:YIG single crystals as a Faraday element," Sens. Actuator A-Phys. **119**, 345–348 (2005).
37. Z. Jin, Z. Mics, G. Ma, Z. Cheng, M. Bonn, and D. Turchinovich, "Single-pulse terahertz coherent control of spin resonance in the canted antiferromagnet $YFeO_3$ mediated by dielectric anisotropy," Phys. Rev. B **87**, 094422 (2013).
38. D. Grass, J. Fesel, S. G. Hofer, N. Kiesel, and M. Aspelmeyer, "Optical trapping and control of nanoparticles inside evacuated hollow core photonic crystal fibers," Appl. Phys. Lett. **108**, 221103 (2016).